# Frequency combs in optically-injected terahertz ring quantum cascade lasers

Md Istiak Khan,[1,a)] Zhenyang Xiao,[2,1] Sadhvikas J. Addamane[3,4] and David Burghoff[2,1]

[1] *University of Notre Dame, Electrical Engineering, Notre Dame, Indiana 46556, USA*
[2] *University of Texas at Austin, Electrical and Computer Engineering, Austin, Texas 78712, USA*
[3] *Center for Integrated Nanotechnologies, Sandia National Laboratories, Albuquerque, New Mexico 87123, USA*
[4] *Sandia National Laboratories, Albuquerque, New Mexico 87185, USA*
[a)] *Author to whom correspondence should be addressed: mkhan6@nd.edu*

Quantum cascade lasers (QCLs) have emerged as promising candidates for generating chip-scale frequency combs in mid-infrared and terahertz wavelengths. In this work, we demonstrate frequency comb formation in ring terahertz QCLs using the injection of light from a distributed feedback (DFB) laser. The DFB design frequency is chosen to match the modes of the ring cavity (near 3.3 THz), and light from the DFB is injected into the ring QCL via a bus waveguide. By controlling the power and frequency of the optical injection, we show experimentally and theoretically that combs can be selectively formed and controlled in the ring cavity. The potential for soliton generation and efficient power extraction through the bus waveguide presents exciting opportunities for integrating this technology in compact comb designs.

## I. INTRODUCTION

Integrated chip-scale frequency combs have diverse applications, including communication[1], frequency synthesis[2], optical ranging[3], and gas spectroscopy[4–8]. In the terahertz (THz) and mid-infrared (MIR) regions, the primary compact source for generating frequency combs is the quantum cascade laser (QCL)[9–11]. While QCLs do not easily form pulses due to their short gain recovery time, other kinds of combs, such as frequency-modulated (FM) combs, still readily form[11,12]. Within a Fabry-Perot QCL cavity, fast gain dynamics lead to strong spatial hole burning[13,14], and the asymmetry of the field at the facets gives rise to phase potentials. This nonlinearity is balanced by the cavity dispersion, and this balance enables the generation of phase-coherent FM combs[14–16].

An alternative method for generating frequency combs involves the use of an integrated QCL ring, which could potentially offer additional advantages due to the lack of strong coupling between fields traveling in counterpropagating directions. Additionally, rings' integrated natures and potential for dissipative Kerr soliton generation make them particularly intriguing. Dissipative Kerr solitons are ultra-short pulses capable of propagating through a dispersive, nonlinear medium while retaining their shape and amplitude[17–19]. The generation of solitons relies on a delicate balance between dispersion, Kerr nonlinearity, parametric gain, and cavity loss[17]. Recently, THz QCL ring frequency combs have been demonstrated using both defect[13] and defect-free cavities[20]. In a cavity with defects, the ring resonator still produces spatial hole burning (SHB) due to a reflection point within the defect. On the other hand, in a defect-free cavity, the absence of SHB, combined with anomalous group velocity dispersion, enables the generation of solitons in ring QCL microresonators[21].

All these configurations utilize the ring QCL by itself without the need for external pumps. This letter discusses another approach, incorporating an integrated external source on the chip. We demonstrate that by injecting the microresonator externally, broadening of the ring spectra is observed, which is consistent with comb formation. Additionally, a bus waveguide is included to facilitate light coupling in and out of the microresonator, resulting in improved power extraction. This design topology also has the possibility of generating solitons, using schemes similar to what has been employed in the mid-infrared regime[17,22–26].

## II. Methods

Fig. 1(a) provides an overview of the fabricated device structure, consisting of three key components: an integrated external pump in the form of a distributed feedback (DFB) laser, a bus responsible for coupling with the ring QCL, and the ring itself. All three parts are fabricated on the same QCL gain medium, a GaAs/AlGaAs heterostructure, which is designed to offer gain within the frequency range of 2.5 THz to 4 THz. The DFB laser is biased above its threshold to ensure efficient coupling into the bus and effective pumping of the microresonator. In contrast, the bus is biased at or slightly below threshold, allowing them to act as highly transparent waveguides rather than active media. The DFB laser is designed to lase at 3.29 THz and is configured as a first-order DFB[27], ensuring appropriate coupling into the bus. Simulation of the device's eigenmodes was carried out using a commercial finite element method (FEM) solver (COMSOL). The DFB was fabricated with a periodicity of 13.5 µm, with each section measuring 6.75 µm. Fig. 1(b) presents the simulated eigenmodes. The lowest loss eigenmode obtained in our design corresponded to the frequency of 3.29 THz.

Ensuring the proper coupling of light between the bus and the microresonator is of utmost importance. This is achieved through careful design considerations and simulations. Specifically, a consistent 2 µm gap between the distributed feedback (DFB) laser and the bus is maintained, with the coupling validated through FEM simulations. Additionally, coupled mode theory simulations[28] are employed to assess the coupling between the ring and the bus, exploring the impact of different distances on the degree of light coupling into the microresonator. Various configurations are devised to achieve different critical coupling points based on the round-trip loss of the resonator. In the presented study, Figure 1(c) demonstrates the simulation results, indicating critical coupling at a distance of 2.6 µm between the bus and the ring. Notably, in the specific device showcased, the distance is set at 2 µm, resulting in an over-coupling. By appropriately biasing the bus and microresonator, we possess the flexibility to adjust the round-trip loss according to our needs. At threshold, the quality factor (Q) of the resonator becomes essentially infinite due to gain clamping.

In addition to proper coupling, another crucial factor is the resonance matching between the DFB laser mode and the mode of the ring. Compared with mid-infrared QCLs, THz QCLs offer much less tuning, as the same change in index results in a lower change in frequency. For these DFB lasers, only several hundred MHz of tuning was available. To ensure this resonance aligned and effectively coupled into the microresonator, we fabricated multiple devices with slight variations in the ring radius. Experimental evidence demonstrated successful light coupling into the microresonator when the modes were matched appropriately. However, one downside of this approach was that rings have large areas when compared with Fabry-Perot lasers, so sweeping the ring radii consumed a significant amount of area on the chip. This necessitated the use of lower-diameter devices. This makes direct detection of the beatnote difficult in these structures.

## III. Results

In our device, when the DFB laser is activated and the bus is maintained slightly below its threshold, we observe a comb-like broadening of the ring spectra. We decided to experimentally verify the proper coupling of light into the ring. To this end, we conducted tuning experiments by varying the ring bias while keeping the DFB bias constant. Throughout the measurements, the bus was held at threshold, and the light was collected from the output facet of the bus. The presence of a resonant dip in transmission through the bus indicates successful light coupling into the resonator, as shown in Figure 2(a). We repeat this experiment for different DFB biases and normalize the output by dividing it by the DFB laser power when the ring is off. The results reveal a decrease in coupled light as the DFB bias increases.

We also perform a similar experiment on another device where the modes are further apart. In this case, no dip in transmission is observed, except for minor variations within the experimental noise range for different DFB biases. Figure 2(c) displays the spectrum of the DFB mode and the ring modes, obtained individually with the other section turned off while the bus is biased. The modes exhibit a relatively close separation of 15 GHz, with the DFB peak at 3.284 THz and the ring peak at 3.269 THz. However, we were unable to resolve each peak when both were on. To better understand the difference between the mode lines, we examine the heterodyne beatnote between the DFB and ring, as measured using a bias tee off the ring. Figure 2(d) presents the beatnote map, where we vary the DFB bias while keeping the ring bias constant at 104 mA. As evidenced by the beatnote frequencies, the modes progressively separate as the DFB bias current increases. This observation is consistent with the transmission data, which shows reduced light collection from the bus output at higher DFB biases.

Importantly, our experimental results reveal a clear trend of increasing separation between the DFB and ring modes as the DFB bias current is tuned to higher values. This trend is supported by both the beatnote frequencies and the transmission data, indicating reduced light coupling into the bus output with higher DFB biases. It is noteworthy that the beatnote experiment provides a more accurate representation of the mode separation compared to the individual spectrum data, as it allows simultaneous observation of both modes. The observed linear trend in both the beatnote and spectrum experiments further confirms the gradual shift between the two modes.

Figure 3(a) displays the spectrum of the device output. The ring is maintained at a constant bias of 104 mA while the bus is held at threshold. The output spectra are presented as the DFB bias is varied from 30mA to 35mA. Figure 3(b) illustrates the corresponding mode difference between the ring and DFB mode. In the absence of injection, the ring operates in a single mode, as depicted in Figure 2(c). However, in the presence of the pump, we observe a broadening of the ring spectrum. The highest-bandwidth spectrum is obtained when the modes are closest together. As the modes move further apart, a smaller number of modes are observed to be lasing.

A common method to verify the formation of a comb output is by measuring the beatnote on a fast detector[10,29]. A narrow beatnote and low phase noise usually indicate comb formation, although additional coherence measurements are usually needed to verify this[30]. However, due to the high mode spacing of this design (required by the sweep), we were unable to perform this experiment on the current device. Instead, we used a superresolution approach to analyze the peak data in the spectrum, zero-padding the interferogram and fitting a parabola to each individual peak to determine the position of the mode peaks to an uncertainty smaller than the instrument resolution[31]. While this approach cannot *resolve* peaks that are separated by less than the instrument resolution, it can find the centroid to a precision determined by the signal-to-noise ratio. Figure 4(b) presents the peaks of each mode, and Fig. 4(c) shows the difference frequency between each pair of modes. All of the difference frequencies agree with each other within the error, with the exception of the peak that corresponds to the combined DFB and ring mode. Since the DFB mode and ring modes cannot be resolved in this way, the calculation yields a difference that differs from the others by around 2.5 GHz. This value is essentially a combination of the two signals and is consistent with the observed heterodyne beating.

To gain further insights into the process of comb formation in ring QCLs, we conducted numerical simulations that employed a master equation formalism derived from a two-level model[14,32,33]. In ring QCLs, the presence of ultrafast carrier lifetimes often leads to asymmetric gain in the active region[34]. This asymmetry is effectively modeled by using linewidth enhancement factor (LEF), α, which was found to be necessary to achieve non-continuous wave radiation[20]. In Figure 4(d), the power spectrum is depicted after 10,000 roundtrips. In the model, we assume a linewidth enhancement of α = 0.45, a coherence lifetime of $T_2$ = 170 fs, a population lifetime of $T_1$ = 40 ps, and a peak small-signal gain of 30 cm⁻¹. The ring's radius measures 300 μm. We assume that 70 percent of the light exits the ring through the bus each roundtrip, which is consistent with our finite element simulations. The injection of light is represented in our model as a source term. Figure 4(e) illustrates the time trace across five roundtrips. Both experimental results and numerical simulations exhibit a reasonable level of agreement, demonstrating the formation of a frequency comb. In the time domain, the simulation's intensity profile also reveals the emergence of modulation-like structures, providing additional support for the potential of soliton formation using injected ring QCLs.

## IV. Conclusion

In this work, we presented a device designed to investigate comb formation in a ring terahertz quantum cascade laser system. The device consists of three components, all fabricated from the same QCL material. The DFB laser is used to inject a single-mode frequency into the ring QCL, and a long bus waveguide facilitates the coupling of light from the DFB laser to the microresonator and enables efficient light extraction at the output facet. Our results confirmed the broadening of the ring spectrum and the emergence of additional lasing modes upon injection of light from the DFB laser. Although direct beatnote measurements were not feasible in this work, the observed broadening of the ring spectra and the agreement in the difference frequencies between the modes strongly indicate the formation of a frequency comb. The conducted numerical simulations also exhibit moderate agreement with the observed data. Further investigation is warranted to gain deeper insights into the dynamics of comb formation in this type of coupled QCL system. Future work will focus on optimizing the device design to increase the DFB tuning range and reduce the need for sweeping, allowing for smaller free spectral ranges and direct measurements of the intermode beatnote. Additionally, there is the potential for generating solitons in this structure by detuning the DFB laser, similar to microresonators in the mid-infrared regime[17,22–26]. The generation of solitons and combs in this type of structure offers the advantage of efficient power extraction through the bus waveguide and the potential for integration in chip-scale platforms for more complex designs.

## Acknowledgments

This work was supported by Air Force Office of Scientific Research (AFOSR) grant no. FA9550-20-1-0192. This work was performed, in part, at the Center for Integrated Nanotechnologies, an Office of Science User Facility operated for the US Department of Energy (DOE) Office of Science. Sandia National Laboratories is a multimission laboratory managed and operated by National Technology & Engineering Solutions of Sandia, LLC, a wholly-owned subsidiary of Honeywell International, Inc., for the US DOE's National Nuclear Security Administration under contract DE-NA-0003525. The views expressed in the article do not necessarily represent the views of the US DOE or the United States Government.

## Author Declarations

## Conflicts of Interest

The authors declare no conflicts.

## Author Contributions

**Md Istiak Khan**: Formal analysis (lead); Investigation (lead); Methodology (lead); Fabrication (equal); Software (equal); Writing – original draft (Lead); Writing – review & editing (equal).

**Zhenyang Xiao**: Investigation (supporting); Methodology (supporting). **Sadhvikas J. Addamane:** Fabrication (equal). **David Burghoff**: Conceptualization (lead); Supervision (lead); Software (equal); Writing – review & editing (equal).

## Data Availability

The data that support the findings of this study are available from the corresponding author upon reasonable request.

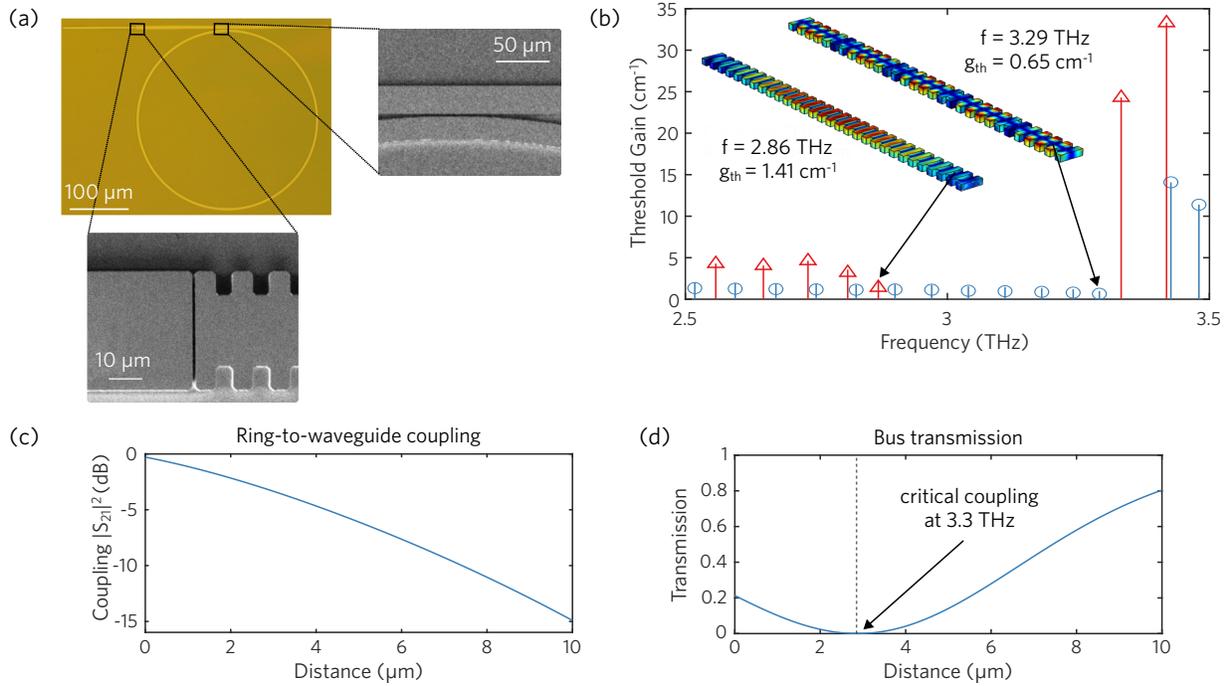

Figure 1 (a) Optical image of the device, with insets showing close-up scanning electron microscopy (SEM) images of the structure. (b) Eigenmode analysis results of the distributed feedback (DFB) laser showing simulated mode profiles of both even and odd modes. The lowest-threshold mode appears at 3.29 THz. (c) Variation of the ring-to-waveguide coupling (as the distance between the bus and the ring is adjusted. (d) Corresponding transmission of a field at resonance. Critical coupling at 3.3 THz occurs at a distance of 3 μm.

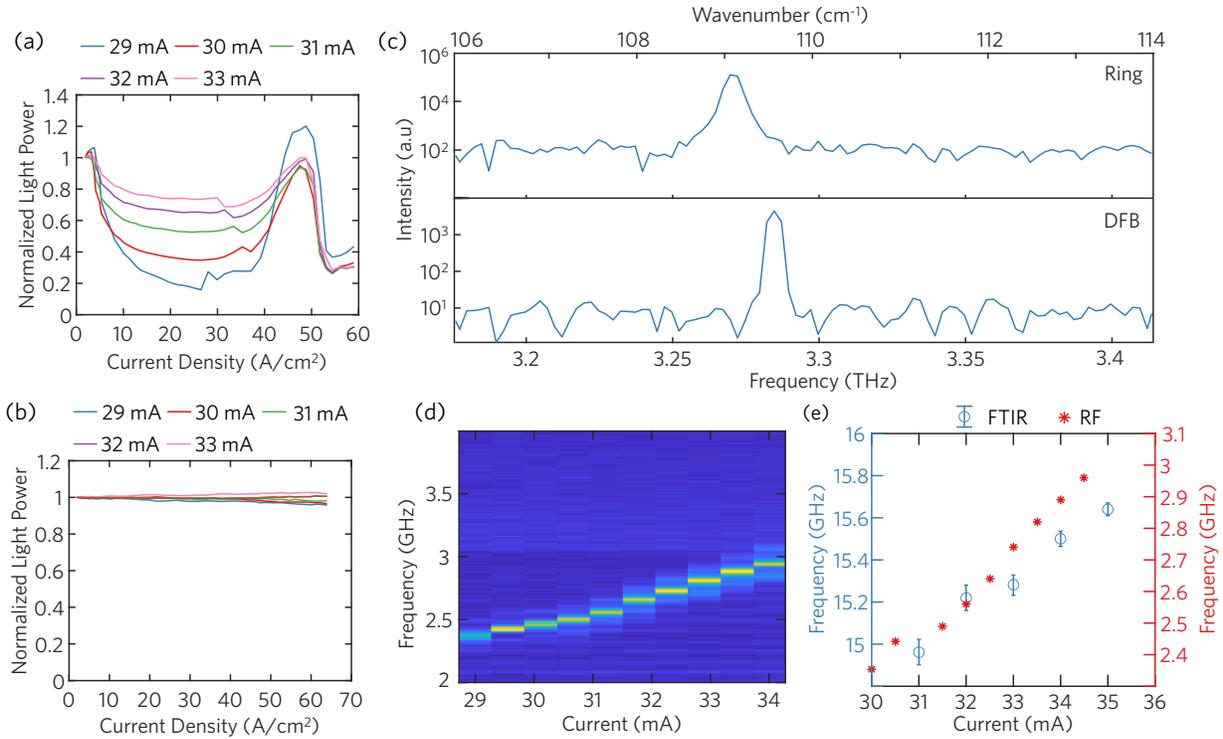

Figure 2 (a) Normalized output light power collected from the bus as the ring bias is swept up to its threshold. Each curve corresponds to different DFB biases, while the bus is held constant at slightly below its threshold. The DFB and ring modes are matched in this device, leading to a noticeable dip in transmission through the bus. (b) Same experiment conducted on a different device, where the modes of the DFB and the ring are further apart. Unlike in Figure 2(a), no dip in transmission through the bus is observed. This indicates that the coupling between the DFB and ring modes is not as effective in this device due to the larger separation between the modes. (c) Spectrum of the DFB and ring individually (i.e., when the other device is not on). A difference of 15 GHz is observed. (d) Beatnote map between the DFB mode and ring mode as the DFB is swept. (e) Tuning of the observed DFB laser frequency, both from the heterodyne beatnote and from the optical spectrum. Both show a similar trend with a similar tuning coefficient.

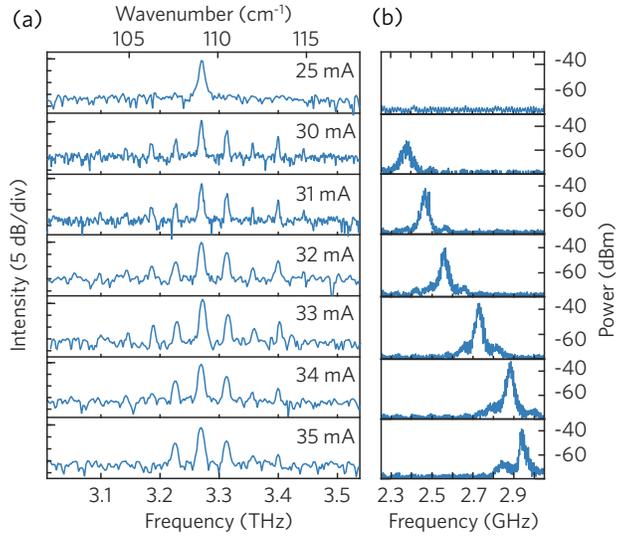

Figure 3 (a) Output spectrum of the ring as the DFB is turned on and swept. The ring spectrum broadens as DFB light is injected, with additional change occurring as the DFB bias is swept. (b) Corresponding beatnote between the DFB and ring modes. As the DFB is detuned from the ring, fewer modes are generated.

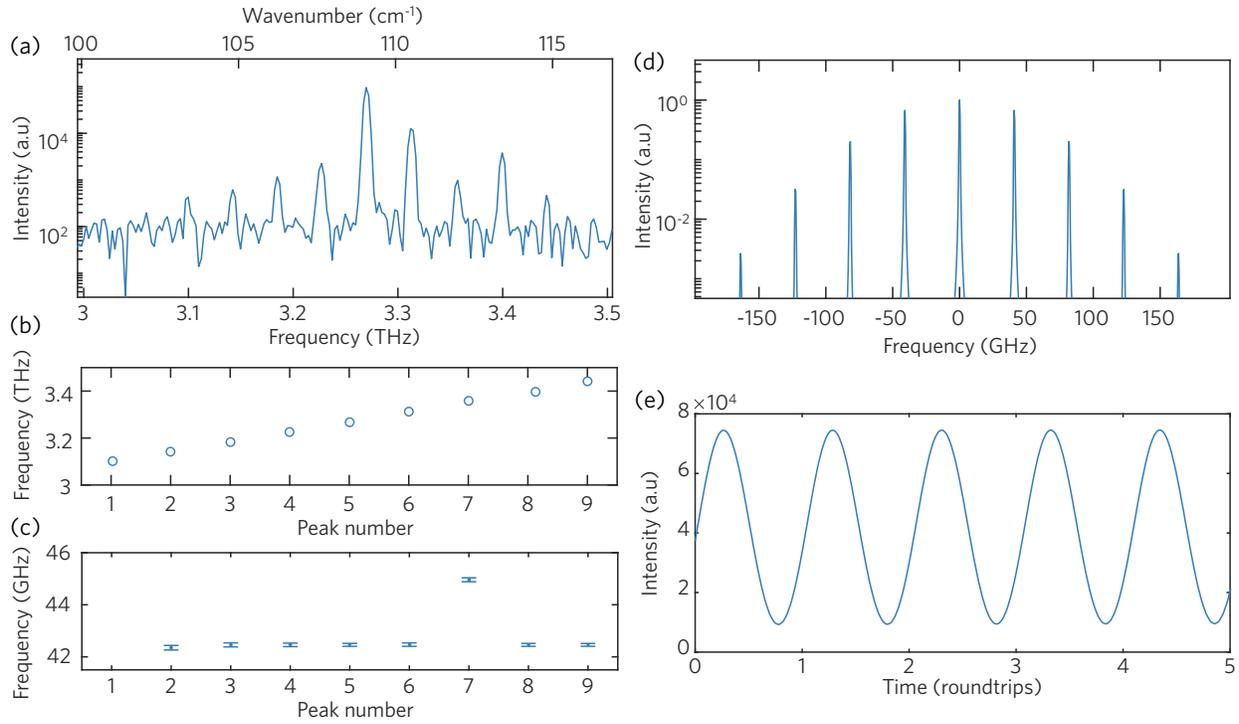

Figure 4 (a) Output spectrum when the DFB is biased at 30 mA and the ring is biased at 104 mA. (b) Observed peaks of each mode in the spectrum and (c) the difference frequency between each successive mode. The difference frequencies are the same except at the injected mode. (d) Numerical simulation of the injected ring QCL, showing the spectrum of the injected ring QCL after 10,000 roundtrips. (e) Simulated laser intensity in the time domain, exhibiting strong intensity self-modulation.